# Coming of Age of the Standard Model

Roger Blandford[1], Jo Dunkley[2], Carlos Frenk[3], Ofer Lahav[4] and Alice Shapley[5].

[1]Kavli Institute for Particle Astrophysics and Cosmology, Stanford University, Stanford, CA, USA.
[2]Department of Physics and Department of Astrophysical Sciences, Princeton University, Princeton, NJ, USA. [3]Institute for Computational Cosmology, Department of Physics, Durham University, Durham, UK. [4]Department of Physics and Astronomy, University College London, London, UK. [5]Department of Physics and Astronomy, University of California Los Angeles, Los Angeles, CA, USA.

Cosmology now has a standard model – a remarkably simple description of the universe, its contents and its history. A symposium held last September in Cambridge, UK, gave this model a 'health check' and discussed fascinating questions that lie beyond it.

The sardonic aphorism, "cosmologists are often in error though seldom in doubt", attributed to Lev Landau, was fair comment when it was coined, at a time when cosmology was still striving to displace philosophical and theological disputation with measurement and calculation. Today, nothing could be further from the truth.

It has been a long voyage. The discovery of quasars in 1963 demonstrated that the universe evolved and eventually led to validation of the general theory of relativity. Even more significantly, the detection of the cosmic microwave background (CMB), in 1964, showed that the cosmos expanded from a hot, dense initial state. Since then, observation and measurement, performed throughout the entire, seventy octave, electromagnetic spectrum are all, essentially, consistent with a description that can be captured in simple assumptions, uncontroversial physics and five basic numbers (age, densities of baryons and dark matter, amplitude and slope of the fluctuation spectrum; M. Kamionkowski), an outcome that would surely have surprised Landau as it did most of the actual cosmologists whose skill and industry have led to its formulation. In modern cosmology, physical quantities are measured with small error and, instead of doubt, there are well-posed new questions, which ought to be addressable.

The meeting, in honour of the first decade of the Kavli Institute for Cosmology, Cambridge, UK, was located at an appropriate point in spacetime for a comprehensive assessment of cosmology. Cambridge was the site of early formulations of evolutionary cosmology and, notably, its antithesis, the steady-state theory. Today it hosts one of the main centers for analyzing data from the Planck satellite (G. Efstathiou) which, together with WMAP (D. Spergel), has measured cosmological quantities with accuracy and confidence. The year 2019 is also a great time to take stock because, while the standard model is robust, measurements of some of the five basic numbers using different techniques disagree by more than the stated uncertainty, pointing to under-appreciated systematics, new astrophysics or, most interestingly, new components, such as non-standard neutrinos (S. Vagnozzi) , fuzzy dark matter or additional bursts of cosmic acceleration to add to the standard model.

This description of the expanding universe requires that disconnected parts of the universe were once in contact. This requires an epoch of inflation (A. Linde, M. Amin) a paradigm, proposed in 1980, which can also account for the observations that the spatial geometry is nearly flat and that the primordial fluctuations share a common entropy and are described by a simple power law that represents small and large scales almost equally. In addition, inflation motivates the quest for gravitational wave modes that should be accessible to the next generation of CMB telescopes including LiteBird from space and the BICEP (Background Imaging of Cosmic Extragalactic Polarization) Array, Simons Observatory and CMB Stage 4 on the ground (C. Pryke, E. Linder).

Of course, cosmology is about much more than agreeing on a few numbers. There is a rich narrative history to be uncovered regarding galaxies and their contents and the key physical processes that govern them. Observations of the most distant galaxies and quasars (R. Ellis, X. Fan) are establishing when ``cosmic dawn'' (A. Fialkov) occurred as primordial hydrogen in the universe was reionized (P. Madau, M. Shull, C. Mason, L. Keating). The prospects for learning much more from the Atacama Large Millimeter /Submillimeter Array (R. Smit, M. Ouchi), GAIA and the James Webb Space Telescope (H. Katz), as well as observations of the 21 cm hydrogen line in emission and absorption (J. Hewitt, P. Sims), are bright. All of this sets the stage for the formation of most of the stars in the Universe – which reaches its peak rate ten billion years ago during "cosmic noon" - when galaxies were roughly three times closer together than they are today and then declined at a later time (S. White).

These galaxies, like the stars they contained, evolved and this evolution can be chronicled by careful observation (R. Genzel, A. Shapley). Of note, there has been remarkable recent progress not only in imaging but also in the acquisition of the much more challenging spectroscopic measures of the stellar and gas content of galaxies over most of cosmic time using state-of-the-art instrumentation. From the theoretical point of view, the evolution of galaxies is described by large cosmological simulations that follow the growth of primordial seeds of quantum origin laid down during inflation and predict the statistical behavior of the assembly of substructure in an organized fashion (R. Wechsler). Likewise, the collection of dark matter halos that contain galaxies into larger associations such as groups and clusters is well described both in simulations and in astronomical data (S. Allen, S. Bocquet, D. Barnes).

A broad theme in the meeting was the fusion of simulation and observation. There have been advances in the former with the focus shifting from the relatively well-defined emphasis on the gravitational interaction of dark matter to the much richer and less well-characterized treatment of the astrophysics of star and massive black hole formation and their environmental consequences (V. Springel, A. Kravstov, A. Dekel, M. Bourne). In particular, an outstanding goal of galaxy formation studies (S. White) is an understanding of the ways in which young stars and massive black holes regulate their growth by feeding energy and momentum into the interstellar and circumgalactic media. This, in turn, determines the observed sizes, luminosities, shapes, and colours of galaxies (V. Semenov, S. Walch, M. Smith). In addition to learning about the galaxies themselves, there is optimism that the large numbers of objects– approaching ten billion – that will be measured with present and future galaxy surveys (W. Percival) will test the precepts and determine the parameters of the standard model beyond what is possible using the CMB data alone (J. Dunkley). (It will be essential to perform spectroscopy on a fraction of these galaxies using instruments such as the recently commissioned Dark Energy Spectroscopic Instrument.)

A very important example of this approach is the hope that the presence of a pure cosmological constant (as opposed to a more general, if less palatable, dark energy with a preferred frame and an equation of state) can be tested using weak gravitational lensing by Euclid and the Large Synoptic Survey Telescope (R. Mandelbaum, S. Joudaki, B. Joachimi). Comparison of the results so far from the Kilo Degree Survey and Dark Energy Survey highlight the challenge involved if cosmological parameters are to be measured to one percent accuracy. Cosmologists have been quick to incorporate new, machine learning-based, data analysis techniques and this will surely be necessary to perform these measurements. A major challenge is to use the observed, luminous galaxies as tracers of the invisible, but dominant dark matter. The approach is to emulate the Universe (H. Peiris, U. Seljak) and use the observational data to "marginalize over the astrophysics" in a self-consistent fashion. Even if we fail to meet this goal, we will still be able to describe and, perhaps, explain the physics behind the births, marriages and degeneration of galaxies. In other words, we should soon learn if galaxies behave more about fundamental physics or phenomenological astronomy.

Another striking feature of the meeting was the inclusion of multi-messenger cosmology (S. Nissanke), most notably through the study of gravitational radiation (A. Sesana, M. Nagathos). Merging binary neutron stars will measure the Hubble constant (D. Holz) and even the more frequent merging black holes (A. Palmese) can be deployed for this task. The Laser Interferometer Space Antenna (LISA) will be especially important for understanding the frequency of binary black holes (M. Campanelli) as well as the sequencing and symbiosis of galactic "chickens" and black hole "eggs". With the advent of PeV neutrino and 100 EeV cosmic ray "astronomy", there are another seventy octaves of combined non-electromagnetic spectra to explore and open new windows to cosmology.

So, despite the much-publicized tensions in the measurement of the Hubble constant and the amplitude of density fluctuations today, cosmology has much to celebrate right now with the definition of its robust, standard model whose birth dates back to the early 1980s. However, cosmologists must also confess their ignorance. They do not know the identity of dark matter, they cannot explain why a tiny fraction of baryons should survive the early universe, they do not understand the mechanics of inflation and they cannot account for the cosmological constant. Fortunately, as emphasized in a panel discussion at the end of the meeting, many audacious ideas are in play and complementary, new observatories should address these questions as well as advance our understanding of neutrinos and elucidate the complex interplay of stars, massive black holes and intergalactic gas in promoting and regulating galaxy formation and evolution. Most optimistically, as this well-organized meeting made clear, cosmology still has the capacity to surprise us. Perhaps, today, we should update Landau's aphorism to "Cosmologists are starting to understand their errors and to recognize when bold answers to magnificent questions deserve lashings of doubt."

**Reference**
Conference website: https://www.kicc.cam.ac.uk/events/copy_of_past-events/kicc-10th-anniversary-symposium-1

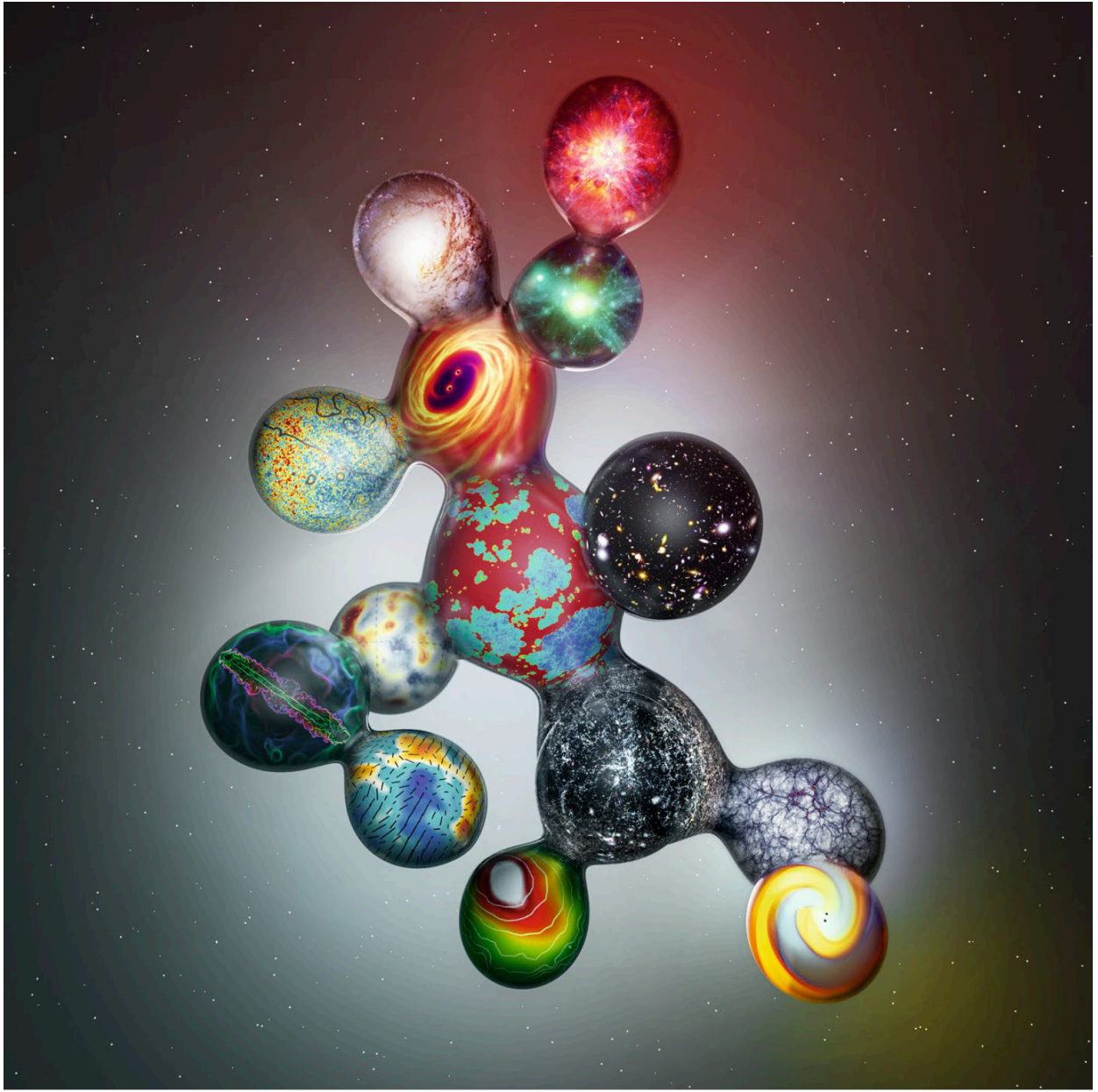

Fig. 1. The multiple fronts on which progress is being made in modern cosmology. This image was used as a poster of the Kavli Institute for Cosmology 10th Anniversary Symposium in Cambridge, UK. Credit: Amanda Smith, Institute of Astronomy. Composite images: Kunesch et al., Millennium XXL simulation, SDSS, Planck, Keating et al., STScI, DES, Bourne et al., Fiacconi et al., FABLE simulation, Illustris simulation, ESA/Hubble, NASA and the LEGUS team.